\documentclass[altaffilletter,aps,nofootinbib,twocolumn,prd,eqsecnum,preprintnumbers,superscriptaddress]{revtex4-1}
\pdfoutput=1
\usepackage[caption=false]{subfig}
\usepackage{graphicx}
\usepackage{amsmath}
\usepackage{amsfonts}
\usepackage{amssymb}
\usepackage{color}
\usepackage{bm}
\usepackage{mathrsfs}
\usepackage{epstopdf}
\usepackage{url}
\usepackage{footnote}
\usepackage{textcomp}
\usepackage{dsfont}
\usepackage{ulem}
\usepackage{hyperref}
\usepackage{enumerate}

\makeatletter
\newcommand*{\rom}[1]{\expandafter\@slowromancap\romannumeral #1@}
\makeatother

\begin{document}

\title{Lessons from Black Hole Quasinormal Modes in Modified Gravity}

\author{Che-Yu Chen}
\email{b97202056@gmail.com}
\affiliation{Institute of Physics, Academia Sinica, Taipei 11529, Taiwan}

\author{Mariam Bouhmadi-L\'opez}
\email{mariam.bouhmadi@ehu.eus}
\affiliation{Department of Physics, University of the Basque Country UPV/EHU, Bilbao 48080, Spain}
\affiliation{IKERBASQUE, Basque Foundation for Science, Bilbao 48011, Spain}

\author{Pisin Chen}
\email{pisinchen@phys.ntu.edu.tw}
\affiliation{Department of Physics and Center for Theoretical Sciences, National Taiwan University, Taipei, Taiwan 10617}
\affiliation{LeCosPA, National Taiwan University, Taipei, Taiwan 10617}
\affiliation{Kavli Institute for Particle Astrophysics and Cosmology, SLAC National Accelerator Laboratory, Stanford University, Stanford, CA 94305, USA}

\begin{abstract}
Quasinormal modes (QNMs) of perturbed black holes have recently gained much interest because of their tight relations with the gravitational wave signals emitted during the post-merger phase of a binary black hole coalescence. One of the intriguing features of these modes is that they respect the no-hair theorem, hence they can be used to test black hole spacetimes and the underlying gravitational theory. In this paper, we exhibit three different aspects of how black hole QNMs could be altered in theories beyond Einstein's general relativity (GR). These aspects are \textit{i}) the direct alterations of QNM spectra as compared with those in GR, \textit{ii}) the violation of  the geometric correspondence between the high-frequency QNMs and the photon geodesics around the black hole, and \textit{iii}) the breaking of the isospectrality between the axial and polar gravitational perturbations. Several examples will be provided in each individual case. The prospects and possible challenges associated with future observations will be also discussed. 
\end{abstract}

\maketitle

\section{Introduction}

The pioneering detection of gravitational waves emitted by binary black hole mergers \cite{Abbott:2016blz,LIGOScientific:2018mvr,Abbott:2020niy} has ushered in an exciting era of gravitational wave astronomy. In light of the rapid development of the observational technology, probing spacetimes under strong gravity and testing gravitational theories using gravitational waves has become an intensive field of research. Although most of the observations so far are consistent with the predictions made in the framework of Einstein's general relativity (GR), there is still much room for improvements in terms of the resolution and accuracy. Therefore, the possibilities that the underlying gravitational theory actually differs from GR should still be taken into account{\footnote{Extending gravitational theories of gravity beyond GR has been motivated by both theoretical and astrophysical reasons. These include the resolution of spacetime singularities, incorporating quantum effects in gravity, and the mysterious dark sectors in the universe. For reviews in modified theories of gravity, we refer the readers to Refs.~\cite{Capozziello:2011et,Nojiri:2017ncd}.}}. These deviations would leave some imprints at the vicinity of black holes and may manifest themselves in the emitted gravitational waves.

Typically, a merger event of a binary black hole system, and its associated gravitational wave signals, can be split into three stages \cite{Abadie:2011kd,Berti:2015itd}. The first one is the inspiral stage, during which the separation of the two objects in the binary is much larger than their individual size. In this stage, the two black holes are rotating round each other, with their separation decreasing in time due to the emission of gravitation waves. In the inspiral stage, the system and the gravitational wave waveforms can be modeled by post-Newtonian approaches or effective-one-body formalisms \cite{Pan:2013rra}. The second stage is the merger phase, during which the two black holes merge together. In the merger phase, the gravitational fields are so strong that numerical relativity is commonly required to model the dynamical evolution. The final stage is the ringdown stage, which is essentially the post-merger stage in a coalescence. In this stage, the two initial black holes have merged into a final black hole, while the black hole is still having some distortions. With the emission of gravitational waves, these distortions gradually decay and the final black hole is shepherded to its stationary configuration. In this paper, we are interested in the ringdown stage and will discuss how the gravitational wave signals in this stage can be a promising tool to test gravitational theories.

The gravitational waves emitted during the ringdown phase are characterized by a superposition of decaying oscillations. Because the system is losing its energy by emitting gravitational waves, the frequencies are complex-valued, with the real part describing the oscillation and the imaginary part corresponding to the decay. As opposed to the well-known \textit{normal modes} with everlasting oscillations for a non-dissipative system, these decaying oscillations are dubbed quasinormal modes (QNMs) \cite{Nollert:1999ji} and they base the natural spectrum of a black hole. It turns out that the QNMs and the dynamics during the ringdown stage can be well-described using black hole perturbation theories. More explicitly, one can regard the distortions of the black hole during the ringdown stage as time-dependent perturbations on the stationary background spacetime. The perturbation modes and their dynamics are subject to their associated master equations, from which the QNM spectrum can be determined after imposing appropriate boundary conditions. For the reviews on black hole QNMs, we refer the readers to Refs.~\cite{Kokkotas:1999bd,Berti:2009kk,Konoplya:2011qq}.

One interesting feature of the spectrum of black hole QNMs in GR is that it respects the black hole no-hair theorem. More explicitly, the spectrum is completely determined by the mass, charge, and spin of the black hole. It has nothing to do with what drives the perturbations in the first place. This feature plays a very crucial role in testing gravitational theories using QNMs because if the theory contains additional parameters or other dynamical degrees of freedom, the QNM spectra of the black hole are altered. Therefore, by comparing the QNM spectra of the black holes in modified theories of gravity with those in GR, one can in principle distinguish these gravitational theories from one another.

In this paper, we are going to exhibit three different but equally important features for black hole QNMs, especially focus on the possibilities of using these features to distinguish different theories of gravity. The first feature is based on the aforementioned property of QNMs that they respect the no-hair theorem. In particular, we will take the non-singular black holes in conformal gravity \cite{Chen:2019iuo,Toshmatov:2017bpx} and the charged black holes in the Eddington-inspired-Born-Infeld (EiBI) gravity \cite{Chen:2018vuw} as examples to demonstrate that, due to the existence of the additional parameters in the theory, the QNM frequencies of these black holes are different from their GR counterparts.

The second feature is based on the geometric correspondence between the high-frequency QNMs (with large multipole numbers) and the photon geodesics around the black hole. In the high-frequency limit, or the eikonal limit, most of the astrophysically relevant black hole solutions in GR satisfy the correspondence that the QNM frequencies can be determined by some specific physical quantities defined on the spherical photon orbits. In this paper, we will show explicitly that this correspondence could be violated in some certain classes of modified theories of gravity. In particular, we find that the non-tirvial matter-curvature couplings in the EiBI gravity \cite{Chen:2018vuw}, as well as the non-minimal matter-gravity coupling in the generalized energy-momentum squared gravity (gEMSG) \cite{Chen:2019dip} break the geometric correspondence. Practically, the violation of such correspondence can be a smoking gun for theories beyond GR.

The third feature that we are going to present in this paper is the isospectrality. In GR, the Schwarzschild, Reissner-Nordstr\"om, and Kerr black holes satisfy the isospectrality, namely, the axial and polar gravitational QNMs for each kind of black holes share an identical spectrum \cite{Chandrabook}. The physical origin of the isospectrality is still unclear. However, it has been confirmed that this feature is generically violated in gravitational theories beyond GR \cite{Cardoso:2019mqo}. The isospectrality can be violated through several mechanisms, with the most common ones being by directly changing the background spacetime from their GR counterparts, or by coupling the axial modes or polar modes with additional dynamical degrees of freedom. As a result, the violation of isospectrality can be used as a tool to distinguish modified theories from Einstein's GR. Several examples will be raised to support our argument.

This paper is outlined as follows. In Sec.~\ref{sec.review}, we briefly review the black hole perturbation theory. By taking the Schwarzschild black hole as an example, we present the master equations for the perturbations and the associated boundary conditions. Techniques for calculating QNM frequencies, in particular the semi-analytic Wentzel-Kramers-Brillouin (WKB) methods, are reviewed. In Sec.~\ref{sec.BHMOG}, we present three different aspects of how the QNMs would be altered in different theories of gravity. These aspects include direct comparison of QNM spectra (Sec.~\ref{sec.qnm1}), the geometric correspondence in the eikonal limit (Sec.~\ref{sec.qnm2}), and the isospectrality (Sec.~\ref{sec.qnm3}). We finally conclude in Sec.~\ref{sec.conclu}.

\section{Black hole perturbations and QNMs}\label{sec.review}
In this section, we will briefly review the black hole perturbation theory and the basic concept of QNMs. To exhibit how modified theories of gravity play their role in this field and to compare the results with their GR counterparts, we should start with GR and review some of its well-known results here. After equipping with these basic knowledge, we will go beyond GR and a more detailed discussion on black hole perturbations and QNMs in modified theories of gravity will be provided in the next section.  

In GR, the most two important black hole solutions are the Schwarzschild black hole and the Kerr black hole, which are expected to describe the spacetime of an isolated non-rotating and rotating black hole, respectively. They are both solutions to the vacuum Einstein equations, and they are asymptotically flat. The linear perturbations of these black holes are generically described by their own master equations, and it is these equations that govern the evolution of the perturbation modes. 

Typically, investigations of black hole perturbations can be manipulated in two different but complementary manners. The first one is to consider some small test fields around the black hole, and focus on the evolution of these fields in the spacetime. In this scenario, it is assumed that there is no back reaction from the test fields acting on the background spacetime. Therefore, according to the spin $s$ of these test fields ($s=0,\pm1$ for scalar fields and electromagnetic fields, respectively), the master equations are derived from the linearized equations of motion of these fields in the black hole spacetime, such as the Klein-Gordon equation and Maxwell equations. The second one stands for the gravitational perturbations on the black hole spacetime itself ($s=\pm2$). This type of perturbations is more observationally relevant because the perturbation modes would manifest themselves in the form of gravitational waves. The master equations of the gravitational perturbations are derived by using the linearized Einstein equations. 

\subsection{Master equations: Schwarzschild black hole}

Let us take the perturbations of a Schwarzschild black hole in GR as an example. The most general metric for a static and spherically symmetric spacetime can be written as follows:
\begin{equation}
ds^2=-e^{2\nu(r)}dt^2+e^{2\mu_2(r)}dr^2+r^2d\Omega_2^2\,.\label{sssmetric}
\end{equation}
For a Schwarzschild black hole, the metric functions are given by
\begin{equation}
e^{2\nu}=e^{-2\mu_2}=1-\frac{2M}{r}\,,\label{Schwarzschild}
\end{equation}
where $M$ is the black hole mass. It turns out that for the perturbations $\Psi$ that we are considering here, the master equations can be written in the form of an ordinary differential equation after suitable field redefinitions and separations of variables:
\begin{equation}
\frac{d^2\Psi}{dr_*^2}+\left(\omega^2-V_{p/s}\right)\Psi=0\,,\label{masterSch}
\end{equation}
where $r_*$ is the tortoise radius and $\omega$ denotes the mode frequency. One of the effective potentials, that is, $V_s$, is labeled by the spin $s$ of the test fields. This potential can be expressed in the following unified form:
\begin{equation}
V_s=\left(1-\frac{2M}{r}\right)\left[\frac{l(l+1)}{r^2}+\left(1-s^2\right)\frac{2M}{r^3}\right]\,,\label{reggewhs}
\end{equation}
where $l$ is the multipole number, which naturally appears when decomposing the perturbation modes in a spherically symmetric spacetime. 

It should be emphasized that when $s=\pm2$, the master equation \eqref{masterSch} is the Regge-Wheeler equation \cite{Regge:1957td} and it describes the evolution of the axial gravitational perturbations. This type of gravitational perturbations is also dubbed odd-parity perturbations based on how the modes behave under the parity change. Relative to the axial modes, there is another type of gravitational perturbations, called the polar gravitational perturbations, or the even-parity modes. The master equation of the polar modes for a Schwarzschild black hole in GR, i.e., the Zerilli equation \cite{Zerilli:1970se}, can also be written in the form of \eqref{masterSch}, but with an effective potential $V_p$ as follows:
\begin{equation}
V_p=\frac{2e^{2\nu}\left[\lambda^2\left(\lambda+1\right)r^3+3M\lambda^2r^2+9M^2\lambda r+9M^3\right]}{r^3\left(\lambda r+3M\right)^2}\,,\label{zerriliv}
\end{equation}
where $\lambda\equiv(l+2)(l-1)/2$. Although the effective potentials \eqref{reggewhs} and \eqref{zerriliv} for the axial modes and polar modes look completely different, their master equations are related via a certain transformation \cite{Chandrabook}, As a result, it turns out that the axial modes and polar modes share an identical spectrum. We will come back to this isospectrality issue later.

\subsection{Boundary conditions}
Having obtained the master equations, the task of calculating the QNM frequencies reduces to solving an eigenvalue problem, which is subject to appropriate boundary conditions \cite{Berti:2009kk,Konoplya:2011qq}. For an asymptotically flat black hole spacetime, the effective potential in the master equations is required to vanish at spatial infinity ($r_*\rightarrow\infty$) and at the event horizon ($r_*\rightarrow-\infty$). In the study of QNMs, the master equations can be regarded as wave equations, which are solved by imposing the boundary conditions that only outgoing waves exist at spatial infinity, while only ingoing waves exist on the event horizon. Essentially, the typical shape of the effective potential has a peak close to the photon ring. Therefore, the perturbations can be treated as waves scattering around the peak of the effective potential, and they are subject to the aforementioned boundary conditions.

It should be noticed that the boundary conditions tightly depend on the asymptotic behavior of the black hole spacetime. For a black hole spacetime which is asymptotically de Sitter, the spacetime of interest is bounded by the event horizon and the cosmological horizon. When represented with the tortoise radius $r_*$, they again correspond to $r_*\rightarrow-\infty$ and $r_*\rightarrow\infty$, respectively, where the potential vanishes. Therefore, the boundary conditions for solving the wave equations are the same as the asymptotically flat case. However, if the black hole is asymptotically anti-de Sitter, there is an infinitely high potential barrier at spatial infinity. Therefore, in this case, one can naturally adopt the boundary condition of Dirichlet type when $r_*\rightarrow\infty$ \cite{Berti:2009kk,Konoplya:2011qq,Horowitz:1999jd}.

\subsection{Calculating QNM frequencies}
As we have just mentioned, after deriving the master equations and imposing the proper boundary conditions, one can treat the master equations as an eigenvalue problem, and the QNM frequencies $\omega$ are the corresponding eigenvalues. Due to the complicated expressions for the effective potentials, there is no analytic expression for QNM frequencies, even for the simplest Schwarzschild case. The evaluations of QNM frequencies can be deduced through purely numerical approaches \cite{Leaver:1986gd,Jansen:2017oag}, or by using some semi-analytic methods, such as the one based on the WKB approximation \cite{Konoplya:2019hlu}, and the asymptotic iteration method \cite{Cho:2009cj,Cho:2011sf}. In this subsection, we will briefly review the concept of using the WKB method to calculate QNM frequencies, since this method has been widely adopted in the literature. For a comprehensive review on the WKB approach, we refer the readers to Ref.~\cite{Konoplya:2019hlu} and the references therein.

The WKB method for calculating QNM frequencies is inspired by the resemblance of the problem to the wave-scattering scenario. The method was formulated in the seminal paper \cite{Schutz:1985km}, and then it was improved to higher orders \cite{Iyer:1986np,Konoplya:2003ii,Matyjasek:2017psv,Matyjasek:2019eeu,Hatsuda:2019eoj}. This semi-analytic method turns out to be powerful because, given the effective potential, one can easily calculate the QNM frequencies with reasonable accuracy just by using a single formula. 

The formulation of the WKB method is based on the fact that the problem can be treated as a quantum scattering process through a potential barrier. In QNM scenarios, there are only outgoing and ingoing waves at $r_*\rightarrow\infty$ and $r_*\rightarrow-\infty$, respectively{\footnote{We do not consider the perturbations of black holes which are asymptotically anti-de Sitter here. In these cases, one has to adopt different boundary conditions and the WKB method is not applicable anymore \cite{Konoplya:2019hlu}.}}. Therefore, when viewed from the perspective of scattering processes, there is no incident wave in the system, but the reflected and the transmitted waves have comparable amounts of amplitudes. One assumes that the peak value of the effective potential is slightly larger than $\omega^2$, such that there are two classical turning points near the peak. The solutions far away from the turning points are solved using the WKB approximation up to a desired order, with the proper boundary conditions taken into account. On the other hand, the solutions at the vicinity of the peak are solved by expanding the potential into a Taylor series. Finally, by matching the solutions near the peak with those obtained from the WKB approximation simultaneously at the turning points, the QNM frequencies $\omega$ can be evaluated. Generically, the QNM frequencies are complex-valued, and have discrete spectra. According to the values of the imaginary parts, the QNM frequencies are labeled by the overtone $n$, with the fundamental mode $n=0$ representing the longest-lasting mode.

It should be emphasized that the WKB method is able to give accurate QNM frequencies when the multipole number $l$ is larger than the overtone $n$. This is expected because the validity of the WKB approximation naturally links to the eikonal optics prescription that the wavelength of the scattered waves is much smaller than the length scale of the curvature of the effective potential. This is also why the WKB method has been widely used for the practical purpose because the fundamental modes ($n=0$) have the longest decay time hence are more astrophysically relevant. Later on in Sec.~\ref{sec.qnm2}, we will present an important feature of black hole QNMs, i.e., the geometric correspondence, and show how the similarity between eikonal waves ($l\gg1$) and photons propagating around the black hole can be used to test gravitational theories.

\section{Black hole QNMs in modified theories of gravity}\label{sec.BHMOG}
After reviewing the calculations of QNMs and the associated master equations in GR, it is time to investigate how one can distinguish different theories of gravity from GR using black hole QNMs. More explicitly, in this section, we will focus on three different perspectives to exhibit how the black hole QNMs in GR and some of their features could be changed in different modified theories of gravity.

\subsection{QNM spectrum}\label{sec.qnm1}
As we have mentioned, the QNM spectrum is completely determined by the parameters that characterize the black hole. For example, for a given spin $s$ of the test field, the QNM spectrum of a Schwarzschild black hole is solely determined by the mass of the black hole. If the black hole spacetime, or the underlying theory of gravity, contains additional parameters, then the QNM spectrum would depend on these parameters as well. In this respect, one can directly consider a particular mode, by fixing the multiple $l$ and the overtone $n$, and compare their mode frequencies with those in GR. This turns out to be the most straightforward way of exhibiting how the QNMs would be altered when going beyond GR, and has been widely investigated in the literature \cite{Kobayashi:2012kh,Kobayashi:2014wsa,Blazquez-Salcedo:2016enn,Bhattacharyya:2017tyc,Glampedakis:2017dvb,Chen:2018mkf,Chen:2018vuw,Chen:2019iuo,Moulin:2019ekf,Chen:2020evr,Bouhmadi-Lopez:2020oia,Liu:2020ddo,Cruz:2015bcj,Cruz:2020emz,Liu:2020ola,Flachi:2012nv}. Here, we shall list some examples which follow this direction.

\subsubsection{Non-singular black holes in conformal gravity}
The first example that we will show here is the QNMs of non-singular black holes in conformal gravity \cite{Chen:2019iuo,Bambi:2016wdn,Toshmatov:2017bpx}. One can imagine that when the curvature scale is extremely high, the universe may respect the conformal invariance and it is described by some conformal theories of gravity. In such a conformal phase, the universe is conformally invariant, and therefore, all the black hole solutions which are related via conformal transformations stand on equal footing. It is only after a certain phase transition at a lower curvature scale that one particular solution is selected and remains in our universe. It turns out that the conformal factors can easily regularize the black hole in the sense that the spacetime is everywhere non-singular \cite{Bambi:2016wdn}. 

In Ref.~\cite{Bambi:2016wdn}, by adopting conformal transformations on the Schwarzschild black hole, some non-rotating black hole metrics have been proposed within this framework. Let us consider the following spacetime metric:
\begin{equation}
ds^2=S(r)ds_{\textrm{Schw}}^2\,,\label{conformalBH}
\end{equation}
where $ds_{\textrm{Schw}}^2$ stands for the Schwarzschild line element given by Eqs.~\eqref{sssmetric} and \eqref{Schwarzschild}. The conformal factor $S(r)$ is assumed to be \cite{Bambi:2016wdn}
\begin{equation}
S(r)=\left(1+\frac{L^2}{r^2}\right)^{2N}\,,
\end{equation}
where $N$ is an arbitrary positive integer and $L$ is a new length scale characterizing the underlying conformal gravity. In Ref.~\cite{Bambi:2016wdn}, it was shown that the curvature invariants are well-defined on the whole spacetime and that the spacetime is geodesically complete. Therefore, the spacetime is everywhere non-singular.

Since the metric is different from the Schwarzschild one, it is then natural to ask whether the perturbation equations of the black hole spacetime \eqref{conformalBH} could see the conformal factor $S$. In Ref.~\cite{Toshmatov:2017bpx}, the scalar field perturbations and the electromagnetic perturbations of this black hole have been investigated, assuming that the energy-momentum tensor is covariantly conserved. Furthermore, assuming that the non-singular black hole is governed by the Einstein equations coupled with an anisotropic fluid, the axial gravitational perturbations have been discussed in Ref.~\cite{Chen:2019iuo}. It was found that the master equations of the perturbations mentioned above can be expressed in the form of 
\begin{equation}
\frac{d^2\Psi}{dr_*^2}+\left(\omega^2-V_\textrm{eff}\right)\Psi=0\,,\label{mastereq}
\end{equation}
where the effective potentials for the scalar field, electromagnetic fields, and the axial gravitational perturbations are \cite{Chen:2019iuo,Toshmatov:2017bpx}
\begin{align}
V_{\textrm{eff,s}}(r)=&\,f_{\textrm{Sch}}\left[\frac{l(l+1)}{r^2}+\frac{1}{Z}\frac{d}{dr}\left(f_{\textrm{Sch}}\frac{dZ}{dr}\right)\right]\,,\nonumber\\
V_{\textrm{eff,e}}(r)=&\,f_{\textrm{Sch}}\frac{l(l+1)}{r^2}\,,\nonumber\\
V_{\textrm{eff,g}}(r)=&\,f_{\textrm{Sch}}\left[\frac{l(l+1)}{r^2}-\frac{2}{r^2}-Z\frac{d}{dr}\left(\frac{f_{\textrm{Sch}}\frac{dZ}{dr}}{Z^2}\right)\right]\,,\label{conformapotentials}
\end{align}
respectively, where $f_{\textrm{Sch}}=1-1/r${\footnote{We will rescale all the quantities with respect to the black hole mass by setting $M=1/2$.}}, and $Z\equiv\sqrt{S(r)}r$. One can see that the master equation for the electromagnetic perturbations does not see the conformal factor. The other two kinds of perturbations, i.e., the scalar field perturbation and the axial gravitational perturbations, do differ from their Schwarzschild counterparts, while they reduce to the Schwarzschild results, \eqref{reggewhs}, in the limit $S\rightarrow 1$.

\begin{figure}[h]
\centering
\vspace{5pt}
\includegraphics[width = .4\textwidth]{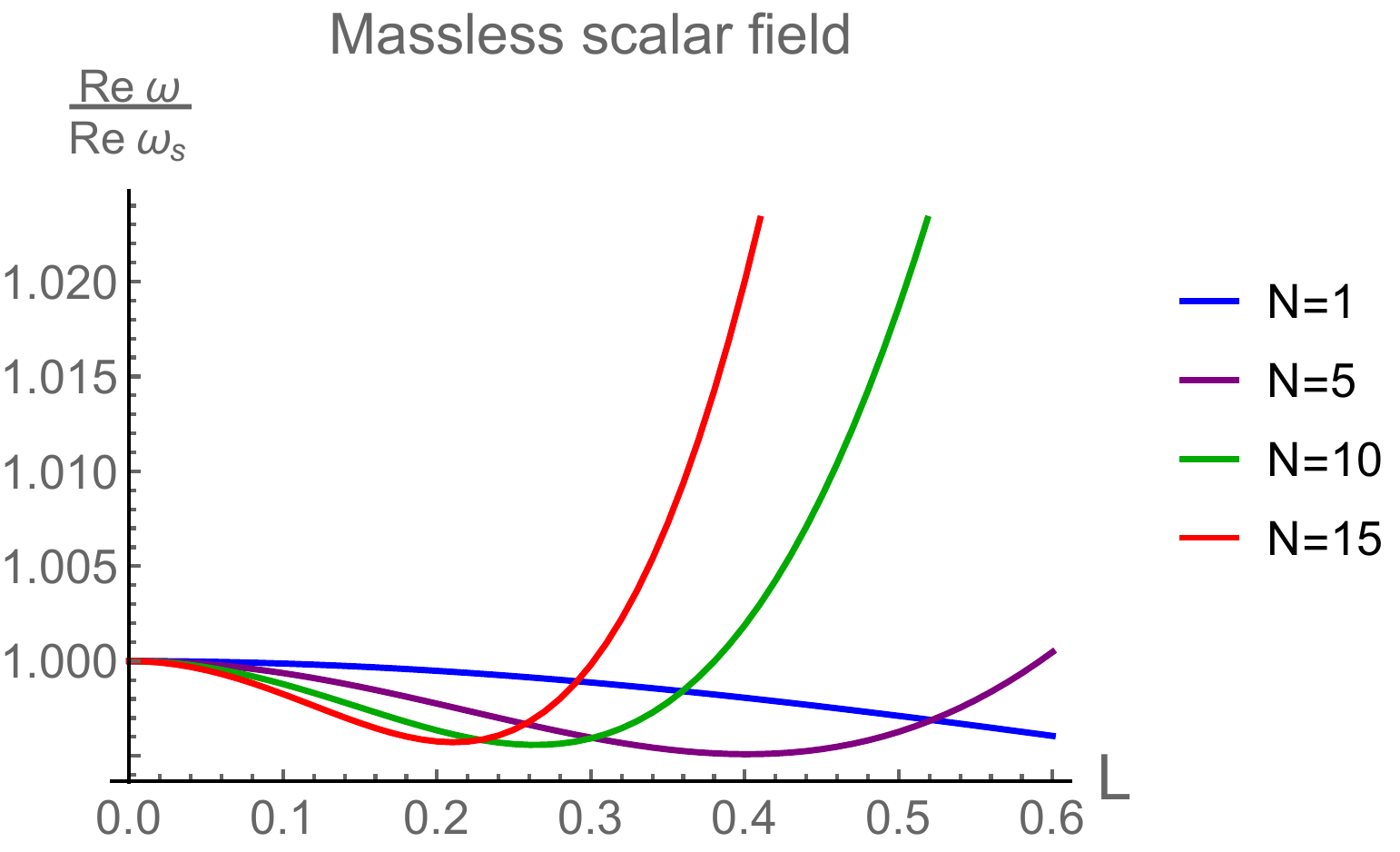}
\includegraphics[width = .4\textwidth]{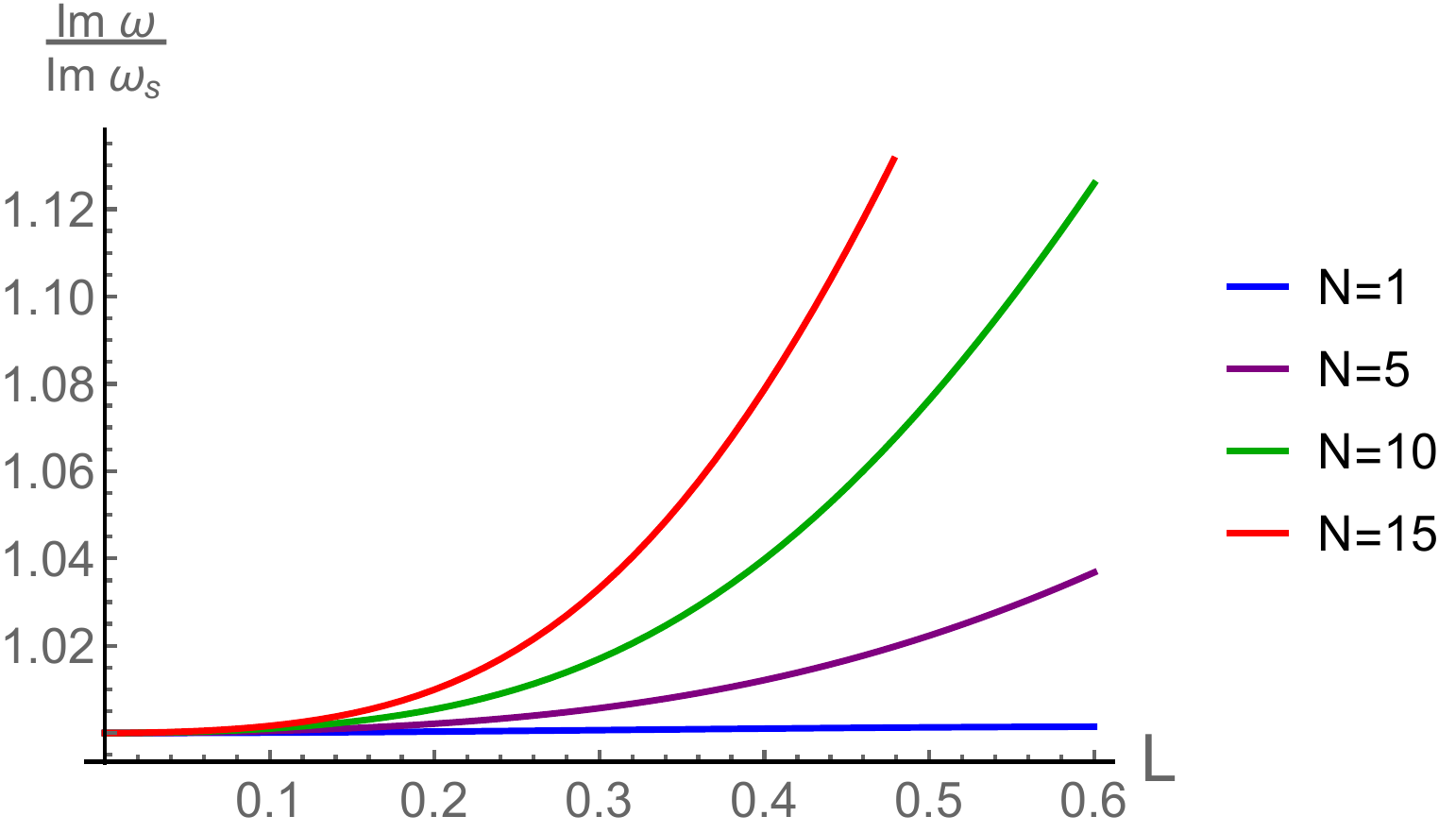}
\vspace{5pt}
\caption{\label{conformalscalar}The real part (upper) and the imaginary part (lower) of the fundamental QNM frequencies for the massless scalar field in the non-singular black hole \eqref{conformalBH} are shown with respect to the length parameter $L$. Different curves represent different values of $N$. The multipole number is chosen to be $l=3$.}
\end{figure}

Using the WKB method mentioned in the previous section, the QNM frequencies can be calculated with high precision. In FIG.~\ref{conformalscalar}, we show the fundamental QNM frequencies ($n=0$) for the massless scalar field of the non-singular black hole \eqref{conformalBH}. It can be seen that the QNM frequencies depend on the conformal factor, and when $L\rightarrow0$ or $N\rightarrow0$, the Schwarzschild results are recovered (see also Figure 3 in Ref.~\cite{Toshmatov:2017bpx}).    

\begin{figure}[h]
\centering
\vspace{5pt}
\includegraphics[width = .4\textwidth]{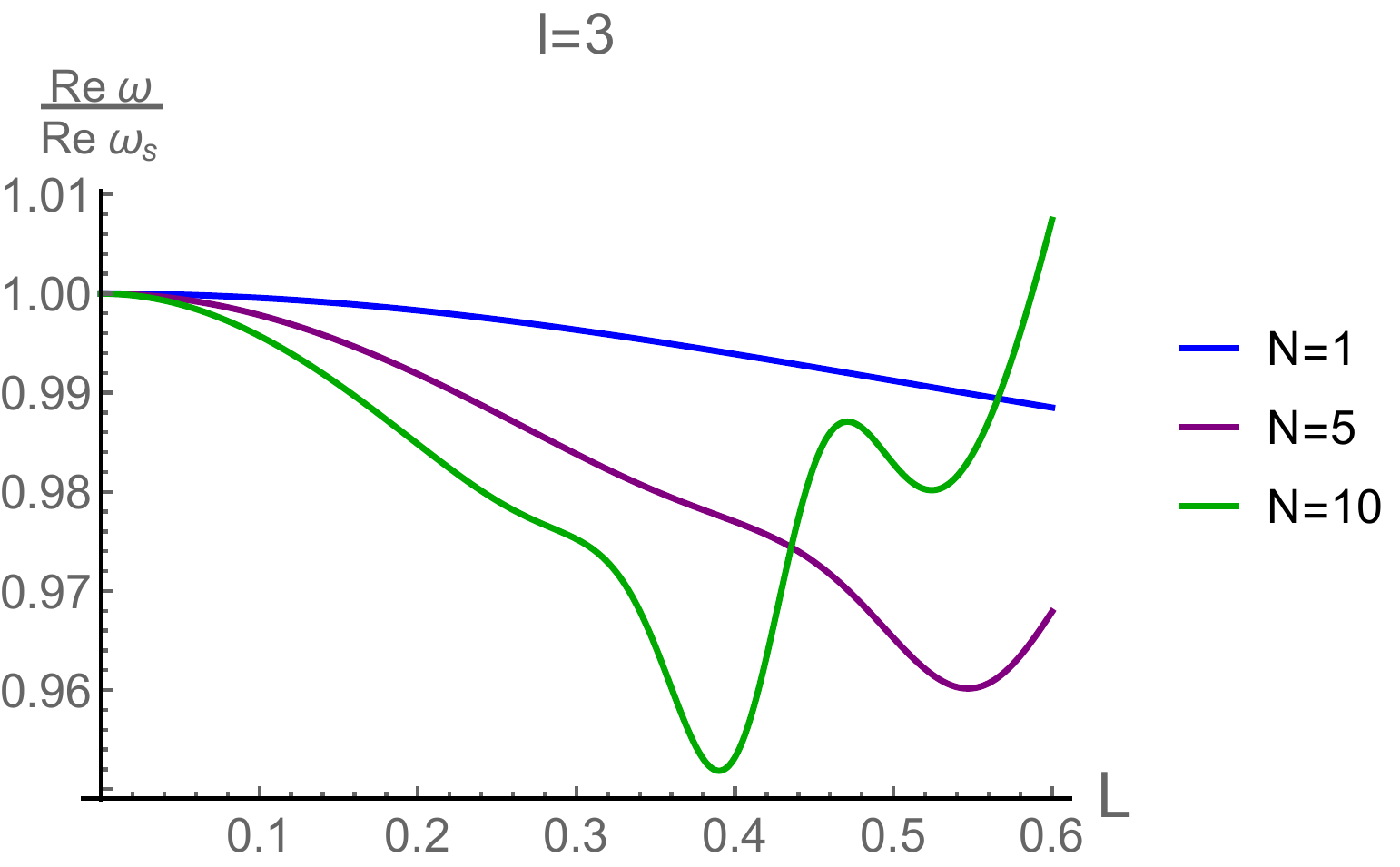}
\includegraphics[width = .4\textwidth]{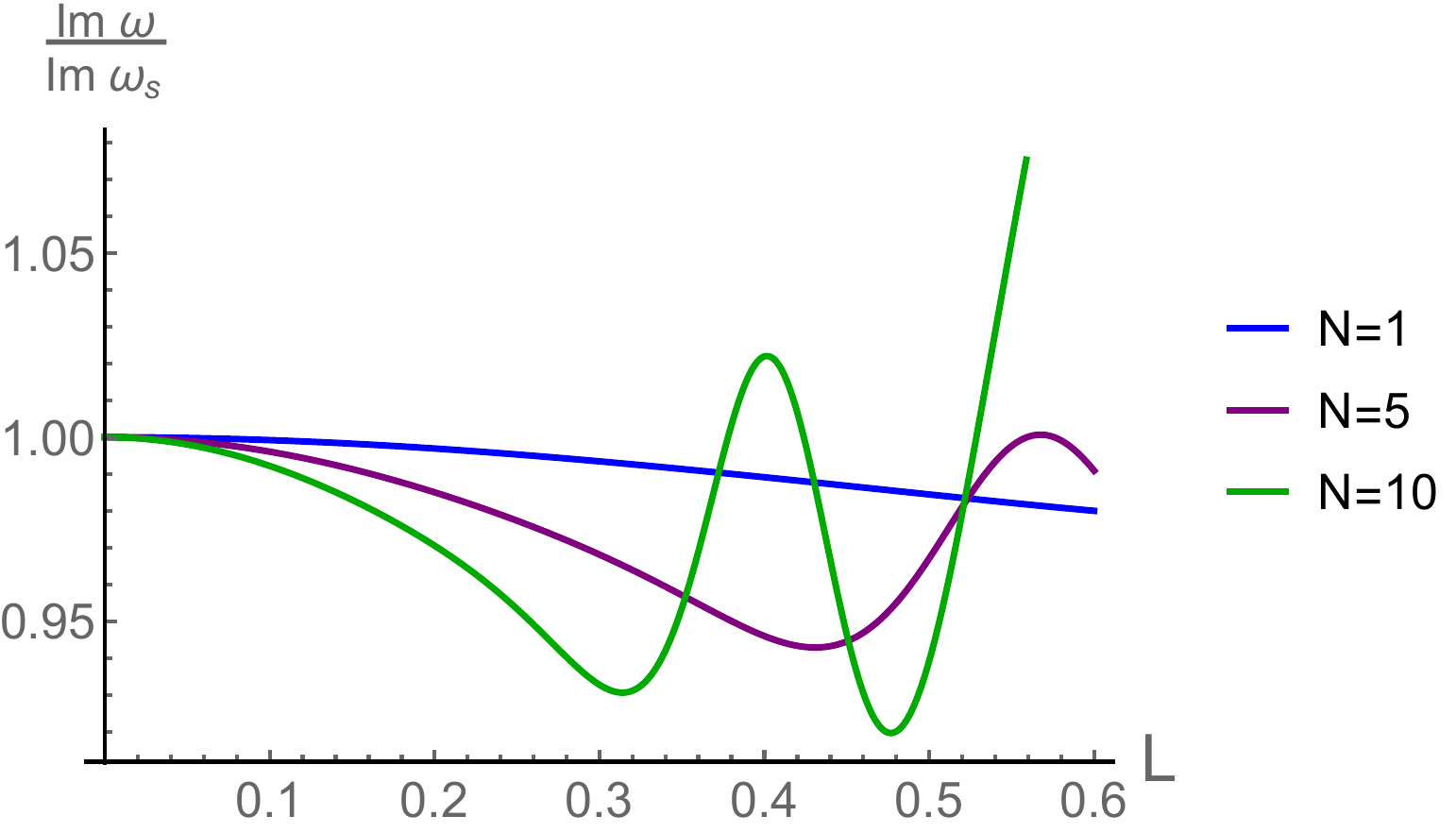}
\vspace{5pt}
\caption{\label{conformalaxial}The real part (upper) and the imaginary part (lower) of the fundamental axial gravitational QNM frequencies for the non-singular black hole \eqref{conformalBH} are presented with respect to the length parameter $L$. Different curves represent different values of the index $N$. In this figure, we choose $l=3$.}
\end{figure}

Furthermore, in FIG.~\ref{conformalaxial}, we show the fundamental QNM frequencies for the axial gravitational perturbations. Again, one can see that the QNM frequencies differ from their Schwarzschild values when $L$ or $N$ increases. As opposed to the QNMs for a massless scalar field, a nontrivial oscillating behavior of the frequencies as a function of $L$ is found \cite{Chen:2019iuo}. Note that we do not show QNMs of the electromagnetic perturbations here because they are not affected by the conformal factor.

\subsubsection{Charged black holes in EiBI gravity}
The second example that we are going to demonstrate the QNM frequencies being altered in modified theories of gravity is the charged black hole perturbations in the EiBI gravity \cite{Banados:2010ix,BeltranJimenez:2017doy}. This particular theory reduces to GR in the absence of matter, while it could largely deviate from GR in the strong gravity regimes in which the energy density of the matter field is high. One of the features of this theory is its ability to ameliorate the big bang singularity in a radiation dominated universe \cite{Banados:2010ix}. In addition, the theory is characterized by a single parameter, and, because of this additional parameter, the spacetime structure will be modified. For instance, the spacetime structure of a charged black hole in the EiBI theory is much different from that of the 
Reissner-Nordstr\"om black hole in GR \cite{Sotani:2014lua,Wei:2014dka}.

We consider the EiBI theory coupled with Maxwell electromagnetic fields. The total action is given by
\begin{equation}
\mathcal{S}=\frac{\epsilon\beta_g^2}{8\pi}\int d^4x\left(\sqrt{\Bigg|g_{\mu\nu}+\frac{R_{(\mu\nu)}}{\epsilon\beta_g^2}\Bigg|}-\sqrt{-g}\right)+\mathcal{S}_m\,,
\end{equation}
where the matter Lagrangian $\mathcal{S}_m$ is described by the Maxwell electromagnetic fields. The Born-Infeld parameter that characterizes the theory is given by $\beta_g$. In the above action, $\epsilon=\pm1$ indicates that the Born-Infeld correction term can be either positive or negative \cite{Banados:2010ix}. It has to be emphasized that the theory is formulated within a metric-affine approach in which the metric $g_{\mu\nu}$ and the affine connection $\Gamma$ are treated independently at the action level. Also, only the symmetric part of the Ricci tensor $R_{(\mu\nu)}(\Gamma)$ is considered in order to respect the projective symmetry.

In this paper, we are going to focus on the axial gravitational perturbations of the charged black holes in the EiBI gravity. The exact solutions for the static and spherically symmetric charged black holes have been obtained in Refs.~\cite{Sotani:2014lua,Wei:2014dka}, and, when expressing the metric in the form of Eq.~\eqref{sssmetric}, we have:
\begin{enumerate}
\item If $\epsilon=+1$:
\begin{align}
e^{2(\nu+\mu_2)}=&\,\frac{r^4}{r^4+r_g^4}\,,\nonumber\\
e^{-2\mu_2}=&\,\frac{r^4+r_g^4}{r^4-r_g^4}\Bigg[1-\frac{r_g^4\beta_g^2}{3r^2}\nonumber\\&-\frac{r}{\sqrt{r^4+r_g^4}}\left(1-\frac{4r_g^4\beta_g^2}{3r}F\left(\frac{1}{4},\frac{1}{2};\frac{5}{4};-\frac{r_g^4}{r^4}\right)\right)\Bigg]\,.
\end{align}
\item If $\epsilon=-1$:
\begin{align}
e^{2(\nu+\mu_2)}=&\,\frac{r^4}{r^4-r_g^4}\,,\nonumber\\
e^{-2\mu_2}=&\,\frac{r^4-r_g^4}{r^4+r_g^4}\Bigg[1-\frac{r}{\sqrt{r^4-r_g^4}}\Bigg(1-\frac{r_g^3\beta_g^2}{3}B\left(\frac{1}{4},\frac{1}{2}\right)\nonumber\\&+\frac{2\sqrt{2}r_g^3\beta_g^2}{3}F\left(\cos^{-1}\frac{r_g}{r},\frac{1}{\sqrt{2}}\right)\Bigg)-\frac{r_g^4\beta_g^2}{3r^2}\Bigg]\,,
\end{align}
\end{enumerate}
where $r_g\equiv\sqrt{Q_*/\beta_g}$ and $Q_*$ stands for the charge of the black hole. We have used $F(..,\,..;\,..;\,..)$, $B(..,\,..)$, and $F(..,\,..)$ to denote the hypergeometric function, the Beta function, and the elliptic function of the first kind, respectively \cite{Abramow}. It can be shown that the above solutions reduce to the Reissner-Nordstr\"om solution when $r\gg r_g$.

In the case of charged black holes, the gravitational fields and the electromagnetic fields are coupled. Therefore, the resulting master equations governing the perturbations turn out to be two coupled equations. The results are \cite{Chen:2018vuw}
\begin{align}
&\frac{d^2H_1}{dr_*^2}+\omega^2H_1=-\frac{4Q_*\lambda e^{2\nu}\sqrt{\sigma_-}}{\sigma_+r^3}H_2\nonumber\\
&+\left[\frac{\bold{S}_{,r_*r_*}}{\bold{S}}+\left(4\lambda^2+2\right)\frac{e^{2\nu}}{r^2}\left(\frac{\sigma_+}{\sigma_-}\right)+\frac{4Q_*^2}{r^4\sigma_+}e^{2\nu}\right]H_1\,,\label{EiBImaster1}\\
&\frac{d^2H_2}{dr_*^2}+\omega^2H_2=-\frac{4Q_*\lambda e^{2\nu}\sqrt{\sigma_-}}{\sigma_+r^3}H_1\nonumber\\
&+\left[-\bold{W}\left(\frac{\bold{W}_{,r_*}}{\bold{W}^2}\right)_{,r_*}+\frac{4e^{2\nu}\lambda^2}{r^2}\left(\frac{\sigma_-}{\sigma_+}\right)\right]H_2\,,\label{EiBImaster2}
\end{align}
where $H_1$ and $H_2$ correspond to the electromagnetic perturbations and the axial gravitational perturbations, respectively. Note that on the above master equations, we have defined
\begin{equation}
\sigma_\pm\equiv1\pm\frac{Q_*^2}{\epsilon\beta_g^2r^4}\,,
\end{equation}
and $\bold{S}\equiv\sqrt{\sigma_-/\sigma_+}$, $\bold{W}\equiv r\sqrt{\sigma_+}$, and the comma here stands for the derivative with respect to $r_*$. 

One can express the coupled master equations in a matrix form as follows
\begin{equation}
\left(\frac{d^2}{dr_*^2}+\omega^2\right)
\begin{bmatrix}
    H_1  \\
    H_2
    \end{bmatrix}=
    \begin{bmatrix}
    V_{11} &V_{12} \\
    V_{21} &V_{22}
    \end{bmatrix}
    \begin{bmatrix}
    H_1  \\
    H_2
    \end{bmatrix}\,,\label{coupledeq1}
\end{equation}
where $V_{ij}$ is given in Eqs.~\eqref{EiBImaster1} and \eqref{EiBImaster2}. In Ref.~\cite{Chen:2018vuw}, it was proven that in the cases of interest where the WKB method is applicable, the eigenvectors of the potential matrix in Eq.~\eqref{coupledeq1} are approximately constant. Therefore, one can approximately diagonalize the potential matrix and decouple the master equations, giving rise to two decoupled master equations for the electromagnetic perturbations and the gravitational perturbations, with potentials $V_1$ and $V_2$, respectively. Furthermore, in the limit where $\beta_g\rightarrow\infty$, the master equations of the Reissner-Nordstr\"om black hole are recovered. If one further assume $Q_*=0$, the master equations reduce to those of the Schwarzschild black hole. More explicitly, the potentials $V_1$ and $V_2$ reduce to $V_s$ given in Eq.~\eqref{reggewhs}, with $s=\pm1$ and $s=\pm2$, respectively. This is expected because $V_1$ governs the evolution of the electromagnetic perturbations, and $V_2$ corresponds to the axial gravitational perturbations.   

\begin{figure*}[t]
\centering
\graphicspath{{fig/}}
\includegraphics[scale=0.48]{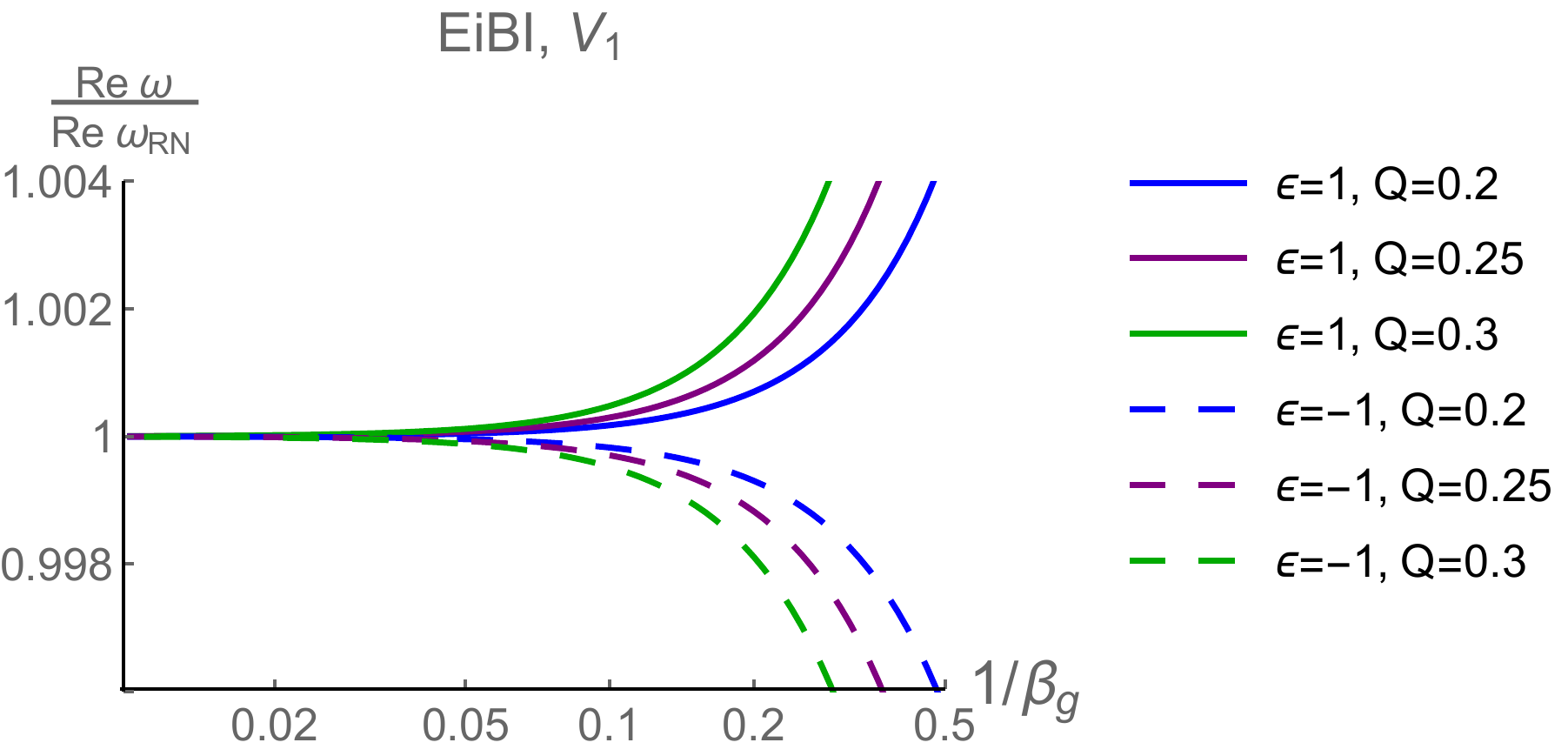}
\includegraphics[scale=0.48]{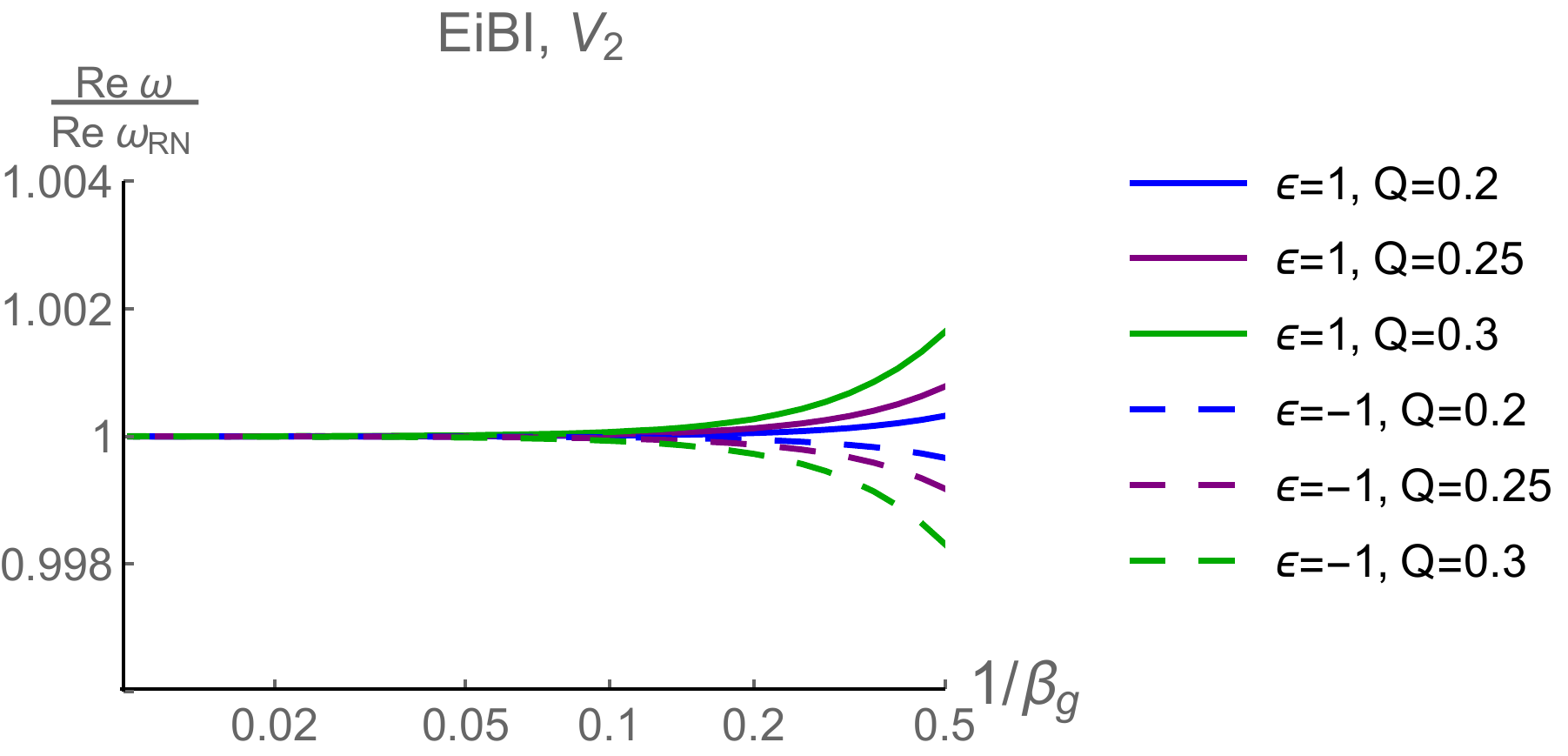}
\includegraphics[scale=0.48]{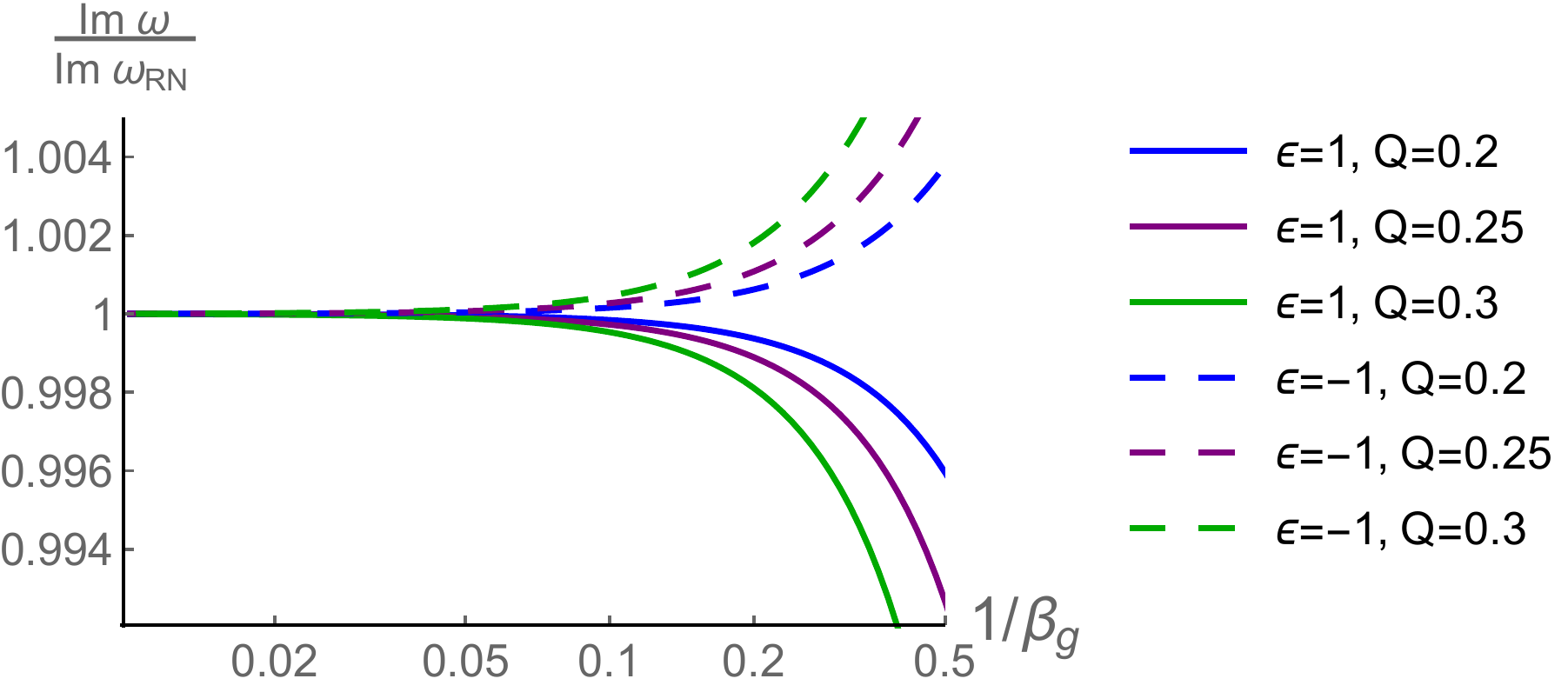}
\includegraphics[scale=0.48]{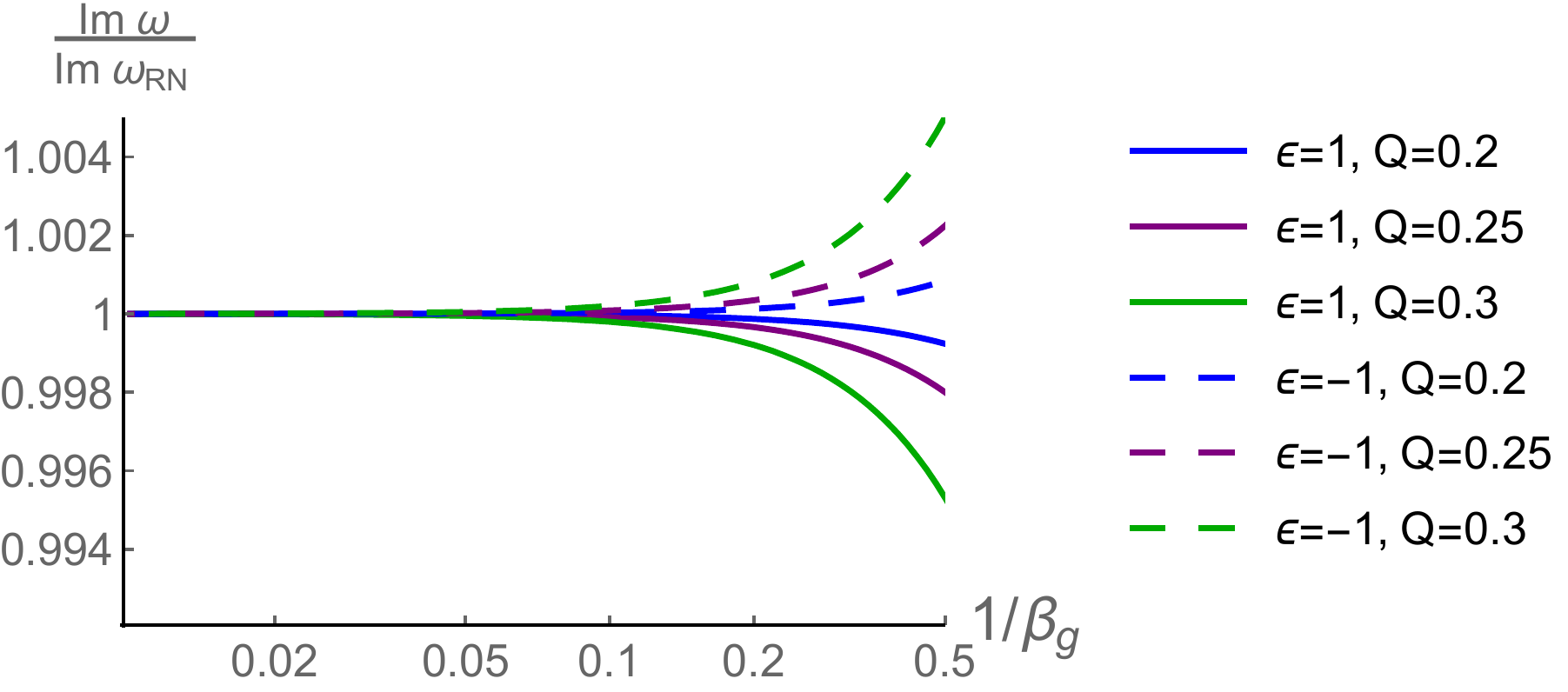}
\caption{The real part (upper) and imaginary part (lower) of the fundamental QNM frequencies of the EiBI charged black holes are presented with respect to $1/\beta_g$. The results are based on the potential $V_1$ (left) and $V_2$ (right), and the multipole number is fixed to $l=2$.} 
\label{fig1}
\end{figure*}

To demonstrate how the QNM frequencies of the EiBI charged black holes are altered by the Born-Infeld coupling constant $\beta_g$, we calculate the fundamental QNM frequencies using the WKB method and show the results in FIG.~\ref{fig1} \cite{Chen:2018vuw}. In this figure, we fix the multipole number to be $l=2$ and consider the decoupled potentials $V_1$ (left) and $V_2$ (right). The solid and the dashed curves correspond to a positive and a negative $\epsilon$, respectively. According to this figure, it can be seen that the QNM frequencies reduce to their GR counterparts when $1/\beta_g\rightarrow0$. In addition, the deviations from the results in GR decrease when the charge is smaller. This is expected as well because the EiBI gravity reduces to GR in the absence of matter fields.  

\subsection{Eikonal QMNs and photon rings}\label{sec.qnm2}

In the previous subsection, we have demonstrated that the QNM spectrum would be generically altered in the presence of additional parameters in theories beyond GR. Testing gravitational theories following this direction thus consists in specifying some particular modes, which should have large enough signal-to-noise ratio, then extracting their mode frequencies and comparing with their GR counterparts. In this subsection, we are going to present another interesting QNM property whose violation would characterize some specific classes of modified theories of gravity. This generic property does not consist in determining the frequencies of specific modes. Instead, it is based on the well-known geometric correspondence between the high-frequency QNMs (eikonal modes with $l\gg1$) and the photon orbits around the black hole \cite{Ferrari:1984zz,Cardoso:2008bp}.

Let us briefly review this geometric correspondence by taking the perturbations of a generic non-rotating black hole spacetime \eqref{sssmetric} as an example. It can been proven that, in the eikonal limit $l\gg1$, the effective potential $V_{p/s}$ in the master equations, regardless of the spin of the test fields, for many black hole solutions in GR can be approximated as
\begin{equation}
V_{p/s}\approx \frac{e^{2\nu}}{r^2}l^2\,,\qquad \textrm{when }l\gg1\,,\label{eikonapo1}
\end{equation}
whose peak is generically located at the photon ring of the black hole \cite{Cardoso:2008bp}. This approximation holds true for the perturbations of the Schwarzschild black holes and the Reissner-Nordstr\"om black holes. As long as the black hole is asymptotically flat, such that the outgoing boundary conditions are valid, one can then use the WKB method mentioned in the previous section to calculate the QNM frequencies. According to Ref.~\cite{Konoplya:2019hlu}, the calculations of the QNM frequencies highly depend on the properties (the value and the derivatives) on the peak of the effective potential. Therefore, it can be expected that the eikonal QNMs are tightly related to the photon ring of the black hole. Indeed, the real part of the eikonal QNM frequency relates to the orbital frequency on the photon ring. On the other hand, the imaginary part corresponds to the Lyapunov exponent, which characterizes the instability of the orbit.

In fact, many black hole solutions in GR share this correspondence, with some known exceptions in which matters of non-linear electrodynamics are introduced \cite{Toshmatov:2018tyo,Toshmatov:2018ell}. In addition, as can be seen from Eqs.~\eqref{conformalBH} and \eqref{conformapotentials}, the non-singular black holes in conformal gravity we mentioned previously satisfy this correspondence. This is expected because in the eikonal limit $l\gg1$, the wavelength of the propagating modes is much smaller than the curvature scale of the effective potential, hence the waves behave like massless particles which are blind to the conformal factor $S(r)$. In fact, this geometric correspondence can be extended to coupled systems \cite{Glampedakis:2019dqh}. It can also be extended to the case of Kerr black holes \cite{Dolan:2010wr,Yang:2012he}, in the sense that the mode frequencies are related to the conserved quantities on spherical photon orbits{\footnote{The real part of the eikonal QNM frequencies can be further related to the apparent size of the black hole shadow, which is directly related to the spherical photon orbits \cite{Jusufi:2019ltj,Jusufi:2020dhz,Cuadros-Melgar:2020kqn}.}}. 

Because the geometric correspondence between eikonal QNMs and photon geodesics turns out to be a generic property for black hole solutions in GR, the violation of it could be a smoking gun of going beyond GR. Naively, this correspondence can be violated if the peak of the effective potential in the eikonal limit is not at the photon ring. By fixing the radius of the photon ring, the violation can happen if the effective potential in the master equation takes the following approximated form
\begin{equation}
V_{\textrm{eff}}\approx \frac{e^{2\nu}A(r)}{r^2}l^2\,,\qquad \textrm{when }l\gg1\,,\label{eikonapo2}
\end{equation} 
where $A(r)$ is a varying function of the radial coordinate. Such a violation has been found in the Einstein-Lovelock theory with a spacetime dimension $D > 4$ \cite{Konoplya:2017lhs,Konoplya:2017wot}. Here, we are going to exhibit, by using two particular examples, that the geometric correspondence can be violated as well in the presence of non-trivial matter-curvature couplings.

\subsubsection{Charged black holes in EiBI gravity}
In the previous subsection, we have investigated the axial perturbations of the charged black holes in the EiBI gravity. Due to the coupling between the electromagnetic fields and gravitational fields, the master equations are coupled together and they are given by Eqs~\eqref{EiBImaster1} and \eqref{EiBImaster2}. In the eikonal limit ($l\gg1$), the effective potentials can be approximated as \cite{Chen:2018vuw}
\begin{align}
V_1\approx\frac{e^{2\nu}}{r^2}\left(\frac{\beta_g^2r^4+Q_*^2}{\beta_g^2r^4-Q_*^2}\right)l^2\,,\\
V_2\approx\frac{e^{2\nu}}{r^2}\left(\frac{\beta_g^2r^4-Q_*^2}{\beta_g^2r^4+Q_*^2}\right)l^2\,.
\end{align}
Note that these approximated expressions are valid for both $\epsilon=\pm1$. Therefore, the presence of the charge $Q_*$ breaks the geometric correspondence because the peaks of the effective potentials are not located at the photon ring anymore. In fact, in the eikonal limit, the electromagnetic perturbations and the gravitational perturbations, which are governed by the potentials $V_1$ and $V_2$ respectively, propagate independently. As we have mentioned, the violation of the geometric correspondence could be a generic property in the modified theories of gravity with non-trivial matter-curvature couplings. Indeed, as has been shown in Refs.~\cite{Pani:2012qb,Delsate:2012ky}, the EiBI theory can be recast into the Einstein frame in which non-trivial matter-geometry couplings naturally appear.

\subsubsection{Charged black holes in gEMSG}
Now, we are going to present another example that non-trivial matter-geometry couplings could break the geometric correspondence. More specifically, we consider a particular class of modified theories of gravity, namely, the gEMSG \cite{Arik:2013sti,Roshan:2016mbt}, and investigate the axial perturbations of charged black holes in this class of theory. The gEMSG is characterized by the presence of non-minimal matter-curvature couplings and it has been shown to have interesting cosmological and astrophysical implications \cite{Roshan:2016mbt,Nari:2018aqs,Akarsu:2018zxl}. The total action of the gEMSG reads
\begin{equation}
\mathcal{S}=\frac{1}{16\pi}\int d^4x \sqrt{-g}f(R,\bold{T}^2)+\mathcal{S}_m\,,
\end{equation}
where $\bold{T}^2\equiv T_{\mu\nu}T^{\mu\nu}$ is the squared contraction of the energy-momentum tensor. The theory is formulated in the metric formalism in which the metric $g_{\mu\nu}$ is compatible with the connection. Because the matter field is non-minimally coupled with gravity, the energy-momentum tensor is not covariantly conserved, and, if the matter field is assumed to be electromagnetic fields, the corresponding Maxwell equations are modified \cite{Roshan:2016mbt}. In Ref.~\cite{Chen:2019dip}, the modified Klein-Gordon equation in the gEMSG has also been obtained when curvature is non-minimally coupled to a scalar field. 

In Ref.~\cite{Roshan:2016mbt}, the charged black holes have been briefly discussed in the gEMSG after assuming a specific functional form of $f(R,\bold{T}^2)$. It has been shown that the charged black holes differ from the standard Reissner-Nordstr\"om one, as expected. Here, we would like to show, as has been concluded in Ref.~\cite{Chen:2019dip} that, due to the presence of non-minimal matter-curvature couplings, the geometric correspondence between the eikonal QNMs and the photon geodesics would be generically broken. Interestingly, this conclusion can be drawn even without specifying the functional form of $f(R,\bold{T}^2)$, nor obtaining the expressions for the background black hole metrics \cite{Chen:2019dip}.

Since we are considering charged black holes in which the electromagnetic and the gravitational fields are coupled, the master equations for the electromagnetic and the gravitational perturbations are coupled as well. The master equations governing the axial perturbations are \cite{Chen:2019dip}
\begin{align}
&\frac{d^2H_1}{dr_*^2}+\omega^2 H_1=-\frac{4Q_*\lambda e^{2\nu}}{\sqrt{f_R\rho_+}r^3}H_2\nonumber\\
&+\left[\frac{1}{2\rho_+^{1/2}}\left(\frac{\rho_{+,r_*}}{\rho_+^{1/2}}\right)_{,r_*}+\left(4\lambda^2+2\right)\frac{e^{2\nu}}{r^2}\bold{\Gamma}+\frac{4Q_*^2e^{2\nu}}{f_Rr^4\rho_+}\right]H_1\,,\label{master1gemsh}\\
&\frac{d^2H_2}{dr_*^2}+\omega^2 H_2=-\frac{4Q_*\lambda e^{2\nu}}{\sqrt{f_R\rho_+}r^3}H_1\nonumber\\
&+\left[-\bold{Z}\left(\frac{\bold{Z}_{,r_*}}{\bold{Z}^2}\right)_{,r_*}+\frac{4e^{2\nu}\lambda^2}{r^2}\right]H_2\,,\label{master2gemsh}
\end{align}
where we have defined $\rho_\pm$ such that it satisfies the following condition
\begin{equation}
\rho_\pm= 1\pm\frac{f_{\bold{T}^2}Q_*^2}{16\pi^2r^4\rho_+^2}\,.
\end{equation}
We have also defined $f_{\bold{T}^2}\equiv\partial f/\partial\bold{T}^2$, $f_{R}\equiv\partial f/\partial R$, $\bold{Z}\equiv\sqrt{f_R}r$, and $\bold{\Gamma}\equiv\rho_-/\rho_+$ on the above expressions \cite{Chen:2019dip}.

Clearly from the master equations \eqref{master1gemsh} and \eqref{master2gemsh} in gEMSG, the geometric correspondence in the eikonal limit is violated. More explicitly, in the eikonal limit ($l\gg1$), the effective potentials in the master equations can be approximated as
\begin{equation}
V_1\approx\frac{e^{2\nu}}{r^2}l^2\bold{\Gamma}\,,\qquad V_2\approx\frac{e^{2\nu}}{r^2}l^2\,.
\end{equation}
Therefore, the peak of the potential $V_2$, which governs the metric perturbations, approaches the photon ring in the eikonal limit, hence the correspondence between the eikonal QNMs and the photon ring still holds. However, the peak of the potential $V_{1}$ is not located at the photon ring due to the $\bold{\Gamma}$ factor. As long as $\bold{\Gamma}\ne1$, the two perturbations $H_1$ and $H_2$ propagate independently, and they do not share the same eikonal frequencies. The geometric correspondence for the electromagnetic perturbations $H_1$ is valid only when $\rho_\pm=1$, that is, either in the absence of charge $Q_*$, or when the non-minimal matter-curvature coupling is turned off ($f_{\bold{T}^2}=0$). 

Here, we have exhibited another example, in addition to the charged black holes in the EiBI gravity, to support our conclusion that the presence of non-trivial or non-minimal matter-curvature couplings in the theory would generically break the correspondence between the eikonal QNMs and the photon ring of the black hole. Any evidence for the breaking of the geometric correspondence would naturally hint toward physics beyond GR. It should be pointed out that to extract the eikonal QNM frequencies in the ringdown signal using the current observational techniques is still very challenging. However, the other side of the correspondence, that is, the photon ring, could be observationally accessible in the near future. In particular, since the black hole shadow is essentially the impact parameter of the photon ring (or of the spherical photon orbits), the eikonal QNM frequencies can be linked with the apparent size of the black hole shadow \cite{Jusufi:2019ltj,Jusufi:2020dhz,Cuadros-Melgar:2020kqn}. Of course, the current resolution of the Event Horizon Telescope \cite{Akiyama:2019cqa} can only access the image of the supermassive black holes with a certain range of angular size, such as that of M87*, while the gravitational waves whose frequencies associated with this type of black holes are still inaccessible. However, it is still worth noting that the geometric correspondence between the eikonal QNMs and the photon ring (or black hole shadows) can be an important tool to test gravity in the upcoming era of gravitational wave astronomy, hence cannot be overlooked.

\subsection{Isospectrality}\label{sec.qnm3}
In addition to the QNM spectrum and the geometric correspondence mentioned in the previous subsections, we will elaborate on another interesting feature of the QNMs in GR, which can also be used to probe black hole spacetimes and modified theories of gravity, that is, the isospectrality. 

As has been mentioned in the previous section, the gravitational perturbations of a black hole can be divided into the axial modes and the polar modes, depending on how they are changed under parity transformations. For the Schwarzschild black hole in GR, the axial and polar modes are governed by the Regge-Wheeler equation \eqref{reggewhs} (with $s=\pm2$) and the Zerilli equation \eqref{zerriliv}, respectively. Although the effective potentials in these two equations look completely different, it is surprising that the QNMs for these two equations share identical spectrum. It was pointed out by Chandrasekhar \cite{Chandrabook}, in particular for Schwarzschild black holes, that the master equations of axial and polar modes satisfy a certain transformation. Although this transformation has recently been identified as a particular subclass of the Darboux transformation \cite{Darboux:1882}, rendering the origin of the isospectrality more mathematically sound \cite{Glampedakis:2017rar}, a satisfactory explanation from a physical point of view is still unclear. In GR, the isospectrality is also valid for the Reissner-Nordstr\"om and the Kerr black hole. While for the Kerr-Newman black hole, the isospectrality has only been confirmed in the slowly rotating limit \cite{Pani:2013ija,Pani:2013wsa}, partially due to the fact that the separability of the master equations in the general case is still an open question. 

Although the isospectrality is a very interesting feature in GR, it is generically violated in modified theories of gravity. That is why this feature can also be used to probe physics beyond GR. For example, assuming that the master equations are sourceless, in Ref.~\cite{Cardoso:2019mqo}, it has been shown using perturbative approaches that if the master equations of the axial or polar modes deviate slightly from their Schwarzschild counterparts, the isospectrality is generically broken and hence it is very fragile. Therefore, in modified theories of gravity whose black hole solutions are different from the Schwarzschild metric, the isospectrality is generically broken.

The other possibility for breaking the isospectrality in modified theories of gravity is through the coupling of the axial modes or polar modes with additional degrees of freedom. In this case, the additional degrees of freedom can be dynamical and source the gravitational polar modes or axial modes, hence the isospectrality could be broken. It should be emphasized that the breaking of isospectrality through the couplings with additional degrees of freedom in the theory can happen in the scenarios where the black hole spacetimes at the background level are indistinguishable from their GR counterparts. Therefore, these solutions cannot be constrained by the observational tests which are implemented on the spacetime metric, but they can be tested through the viability of isospectrality or other methods based on QNM observations. 

The first example of breaking the isospectrality through the coupling with additional degrees of freedom is the metric $f(R)$ gravity. It is well-known that the Schwarzschild metric remains an exact solution in this theory. However, considering the perturbations of the Schwarzschild black hole in $f(R)$ gravity, it turns out that the additional scalar degree of freedom couples to the polar modes, while the axial modes remain intact \cite{Bhattacharyya:2017tyc,Bhattacharyya:2018qbe,Datta:2019npq}. More explicitly, the master equation of the axial modes is again the Regge-Wheeler equation \eqref{masterSch} and \eqref{reggewhs} (with $s=\pm2$), while that of the polar modes is the Zerilli equation with a non-vanishing source term representing the dynamical scalar degree of freedom. Therefore, the isospectrality is naturally broken.

The second example is the dynamical Chern-Simons gravity \cite{Alexander:2009tp}, which, in addition to the Einstein-Hilbert action, includes a direct coupling between a dynamical scalar field and the Pontryagin scalar. By definition, the Pontryagin scalar is zero in a static and spherically symmetric spacetime. However, it is coupled to the axial modes of the metric perturbations. In Refs.~\cite{Cardoso:2009pk,Molina:2010fb,Kimura:2018nxk,Bhattacharyya:2018hsj}, the perturbations of the Schwarzschild black hole in the dynamical Chern-Simons gravity have been investigated. It turns out that the additional scalar degree of freedom is coupled to the axial modes, and it becomes the source term in the Regge-Wheeler equation. The polar modes, on the other hand, remain unchanged and are described by the sourceless Zerilli equation. 

In fact, the breaking of isospectrality also happens in scalar-tensor theories \cite{Kobayashi:2012kh,Kobayashi:2014wsa,Tattersall:2017erk,Tattersall:2018nve} as well as in the Einstein-dilaton-Gauss-Bonnet gravity \cite{Blazquez-Salcedo:2016enn} because of the coupling of the dynamical scalar field. The key question from the observational perspectives is that whether these additional dynamical degrees of freedom are excited or not in reality, and even if they are excited, whether the signal is strong enough to be detected by the near-future detectors. In fact, in some theories of gravity (e.g. the nonlocal gravity \cite{Deser:2007jk,Deser:2019lmm}), whether the dynamical degrees of freedom are excited or not is not determined by the equations of motion, but depends on the choice of boundary conditions \cite{Chen:2021pxd}, i.e., the values of the fields at spatial infinity. If these additional modes are not excited, or their amplitudes are too small, the tests of gravity by examining the excitation of these modes would be extremely challenging. However, we would like to highlight that in addition to directly comparing the QNM spectra and the tests of geometric correspondence, the isospectrality of QNMs can be a promising possibility to test gravitational theories.

\section{Conclusions}\label{sec.conclu}

Black hole perturbations and the associated QNMs are important in light of the recent developments of gravitational wave observations. It is promising to use QNMs to probe the spacetime at the vicinity of a black hole, hence to test whether our current understanding of black hole physics based on GR is correct or not. In this paper, we exhibit three different but equally important aspects of black hole QNMs and investigate how these aspects can help us to distinguish different gravitational theories. 

The first aspect is the most straightforward one. Based on the fact the black hole QNMs respect the no-hair theorem in the sense that the QNM spectrum for black holes in GR is uniquely determined by the mass, charge, and spin, any deviations of QNM spectrum from its GR counterpart caused by other additional parameters can be a smoking gun for theories beyond GR. As an illustration, we consider two particular examples: the perturbations of non-singular black holes which are conformally related to the Schwarzschild one, and the perturbations of the charged black holes in the EiBI gravity. It can be seen that the additional parameters in the theory would alter the QNM frequencies, making them different from their GR counterparts.

The second aspect is based on the geometric correspondence between the eikonal QNMs and the photon geodesics. In the eikonal limit where $l\gg1$, the QNMs are related to the properties of the spherical photon orbits around the black hole. This correspondence can be extended further to the shadow of the black hole, which is also determined by the spherical photon orbits. In some modified theories of gravity, particularly the theories with non-trivial matter-curvature couplings, this geometric correspondence can be generically violated. We take the charged black holes in the EiBI gravity and those in the gEMSG as illustrative examples. It is well-known that the EiBI gravity can be recast into the Einstein frame with non-trivial matter-geometry couplings. On the other hand, the gEMSG is characterized by the non-minimal matter-curvature couplings in the theory. In both cases, the geometric correspondence is violated because of the non-trivial couplings mentioned above.

The third aspect is the isospectrality between the axial and the polar modes of the gravitational perturbations. Most of the black holes in GR, which are astrophysically relevant such as the Schwarzschild, Reissner-Nordstr\"om, and Kerr black holes, share this interesting property. The isospectrality is generically broken in theories beyond GR. Essentially, this can happen, for example, when the metric functions differ from their GR counterparts, or when there are additional dynamical degrees of freedom inherent in the theory. In the latter case, the isospectrality can be violated even when the background spacetimes share the same form as those in GR. The additional degrees of freedom, if they are excited, would source the axial modes or the polar modes, hence breaking the isospectrality. Any evidence for the violation of the isospectrality, or for the excitation of dynamical degrees of freedom other than gravitational fields, will also be a smoking gun for theories beyond GR.

Indeed, none of the above mentioned aspects is an easy way for the future gravitational wave observations to test strong gravity and even to falsify some modified theories of gravity. However, with the rapid development of the techniques, we can expect to have a better understanding through black hole QNMs toward our universe within the era of gravitational wave astronomy. 

\acknowledgments

CYC is supported by Institute of Physics of Academia Sinica. The work of MBL is supported by the Basque Foundation of Science Ikerbasque. She also would like to acknowledge the partial support from the Basque government Grant No. IT956-16 (Spain) and from the project FIS2017-85076-P (MINECO/AEI/FEDER, UE). PC is supported by Ministry of Science and Technology (MOST), Taiwan, through no. 107-2119-M-002-005, Leung Center for Cosmology and Particle Astrophysics (LeCosPA) of National Taiwan University, and Taiwan National Center for Theoretical Sciences (NCTS).

\end{document}